\documentclass[12pt]{article}
\usepackage{graphicx}
\begin{document}
 \noindent {\footnotesize\it Astronomy Letters, 2011 Vol. 37, No. 8, pp. 550--562}

 \noindent
 \begin{tabular}{llllllllllllllllllllllllllllllllllllllllllllll}
 & & & & & & & & & & & & & & & & & & & & & & & & & & & & & & & & & & & & & & \\\hline\hline
 \end{tabular}

 \vskip 1.5cm
 \centerline {\large\bf Searching for Possible Siblings of the Sun from a Common Cluster}
 \centerline {\large\bf Based on Stellar Space Velocities}
 \bigskip
 \centerline {V.V. Bobylev$^{1,2}$, A.T. Bajkova$^1$, A. Myll\"{a}ri$^3$, and M. Valtonen$^4$}
 \bigskip

{\small\it
 $^1$~Pulkovo Astronomical Observatory, Russian Academy of Sciences, Pulkovskoe sh. 65, St. Petersburg,
 196140 Russia; E-mail: vbobylev@gao.spb.ru

 $^2$~Sobolev Astronomical Institute, St. Petersburg State University, Universitetskii pr. 28, Petrodvorets, 198504 Russia

 $^3$~$\AA$bo Akademi University, Turku, Finland

 $^4$~Helsinki Institute of Physics, Helsinki University, Finland
 }

 \bigskip

{\bf Abstract}---We propose a kinematic approach to searching for
the stars that could be formed with the Sun in a common ``parent''
open cluster. The approach consists in preselecting suitable
candidates by the closeness of their space velocities to the solar
velocity and analyzing the parameters of their encounters with the
solar orbit in the past in a time interval comparable to the
lifetime of stars. We consider stars from the Hipparcos catalog
with available radial velocities. The Galactic orbits of stars
have been constructed in the Allen--Santillan potential by taking
into account the perturbations from the spiral density wave. We
show that two stars, HIP~87382 and HIP~47399, are of considerable
interest in our problem. Their orbits oscillate near the solar
orbit with an amplitude of $\approx$250~pc; there are short-term
close encounters to distances $<10$~pc. Both stars have an
evolutionary status and metallicity similar to the solar ones.

 \bigskip
 \bigskip

\section*{INTRODUCTION}

According to present-day observations, stars are formed in groups
or clusters of various strengths. It is highly likely that the Sun
was formed in some ``parent'' open star cluster.

Comparison of the solar chemical composition with the metallicity
distribution in the Galactic disk led to the conclusion that the
Sun was born $\approx$2~kpc closer to the Galactic center relative
to its current position (Wielen et al. 1996). A discussion of this
problem in light of the currently available data can be found in
Acharova et al. (2010). Revealing possible siblings of the Sun
will allow a better understanding of both the mechanisms of radial
migration in the Galactic disk (Minchev and Famaey 2010;
Shevchenko 2010) and the conditions under which the Sun was formed
(Williams 2010). It is hypothesized that the stars born together
retain their chemical homogeneity for a very long time
(Bland-Hawthorn and Freeman 2004; Bland-Hawthorn et al. 2010).

Therefore, as a method of searching for such stars, these authors
suggested analyzing spectroscopic data to determine the abundances
of various elements. In particular, the spectroscopic method has
been successfully used to prove the dynamical origin of the stars
belonging to the Hyades cluster (Pomp\'eia ia et al. 2011).

Obviously, the stars formed together with the Sun are also of
great interest in studying both the dynamical evolution of the
Solar system and the development of terrestrial life (Valtonen et
al. 2009).

The difficulties of revealing such stars in the solar neighborhood
are associated with the Sun's great age ($\approx$4.6~Gyr), the
dynamical evolution of open star clusters (OSCs) during this time,
and a large uncertainty in Galactic parameters. Numerical
simulations of the dynamical evolution of OSCs show (Chumak et al.
2005; Chumak and Rastorguev 2006a, 2006b) that stellar tails
stretched along the Galactic cluster orbits develop in them with
time. The OSC remnants existing in the form of tails must
completely dissolve and mix with the stellar background in a time
$\approx$2~Gyr (K\"{u}pper et al. 2008). According to the
estimates by Portegies Zwart (2009), 10--60 stars from the parent
(for the Sun) OSC containing $\approx10^3$ members can now be in a
solar neighborhood about 100 pc in radius.

The attempts at searching for the Sun's siblings from a common
cluster are known in the literature. For example, Brown et al.
(2010) searched for suitable candidates among the stars of the
Hipparcos catalog (1997) in an updated version (van Leeuwen 2007).
A list of candidates containing six stars was proposed. However,
when the orbital motion of stars was simulated, the influence of
the perturbations from the spiral density wave was disregarded. In
addition, the search was carried out using only the stellar
parallaxes and proper motions (without radial velocities).
Mishurov and Acharova (2010) showed that the influence of the
spiral density wave in a time interval of 4.6 Gyr leads to a
significant dispersal of the members of an initially compact
cluster in both radial and tangential directions. For about a
hundred dispersed stars to be now observable in a solar
neighborhood $\approx$100 pc in radius, the parent cluster must
contain $\approx10^4$ members.

In this paper, we suggest applying a kinematic approach to
searching for the stars that could be formed with the Sun in a
common parent cluster. This approach consists in (1) selecting
suitable candidates by the closeness of their space velocities to
the solar velocity, (2) constructing the stellar and solar orbits
in the Galactic potential including the perturbations from the
spiral density wave, and (3) analyzing the parameters of the
encounters between the stellar orbits and the solar orbit in the
past in a time interval comparable to the lifetime of stars.

\section*{DATA}

The initial set of kinematic data on the stars from the Hipparcos
catalog (1997) that we used is described in detail in Bobylev et
al. (2010). It contains the parallaxes and proper motions of about
35000 stars taken from a revised version of the Hipparcos catalog
(van Leeuwen 2007). The radial velocities from the PCRV catalog
(Gontcharov 2006) are available for each star.

We selected 162 F, G, and K stars with a relative parallax error
$\sigma_\pi/\pi<15\%$ and magnitude of the total stellar space
velocity $(U,V,W)$ relative to the Sun $\sqrt{U^2+V^2+W^2}<8$~km
s$^{-1}$, where 8 km s$^{-1}$ was estimated from a typical random
error in each of the velocities $U,V,W,$ which are $\approx$2~km
s$^{-1}$. Note also that age and metallicity estimates are
available for a significant fraction of the selected stars
(Holmberg et al. 2009).

Remarkably, two of the six candidates from Brown et al. (2010)
entered into our list of 162 stars: HIP~21158 and HIP~99689.

\section*{THE CONSTRUCTION OF ORBITS}

We calculated the stellar and solar orbits by solving the
following system of equations of motion based on a realistic model
of the Galactic gravitational potential (Fernandez et al.
2008):
\begin{equation}
\ddot{\xi}=-\frac{\partial\Phi}{\partial\xi}-\Omega^2_0(R_0-\xi)-2\Omega_0\dot{\eta},
\end{equation}
 $$
\ddot{\eta}=-\frac{\partial\Phi}{\partial\eta}+\Omega^2_0\eta+2\Omega_0\dot{\xi},
 $$$$
\ddot{\zeta}=-\frac{\partial\Phi}{\partial\zeta},
 $$
where $\Phi$ is the Galactic gravitational potential; the
 $(\xi,\eta,\zeta)$ coordinate system with the center at the Sun rotates
around the Galactic center with a constant angular velocity
$\Omega_0,$ with the  $\xi,\eta,$ and $\zeta$ axes being directed
toward the Galactic center, in the direction of Galactic rotation,
and toward the Galactic North Pole, respectively; $R_0$ is the
Galactocentric distance of the Sun.

We used the Allen--Santillan (1991) model Galactic potential. The
system of equations~(1) was solved numerically by the fourth-order
Runge--Kutta method.

\begin{table}[t]
 \caption{Parameters of he model Galactic potential}
\begin{center}
  \label{Tab1}
\begin{tabular}{|c|c|}\hline
 $M_C$ & 606 M$_G$ \\\hline
 $M_D$ & 3690 M$_G$ \\\hline
 $M_H$ & 4615 M$_G$ \\\hline
 $b_C$ & 0.3873 kpc \\\hline
 $a_D$ & 5.3178 kpc \\\hline
 $b_D$ & 0.25 kpc \\\hline
 $a_H$ & 12 kpc \\\hline
\end{tabular}
\end{center}
\end{table}

In the Allen--Santillan (1991) model, the Galactocentric distance
of the Sun is taken to be $R_0 = 8.5$~kpc and the circular
velocity of the Sun around the Galactic center is $V_0 =
|\Omega_0|R_0 = 220$~km s$^{-1}$. The axisymmetric Galactic
potential is represented as the sum of three components---the
central (bulge), disk, and halo ones:
\begin{equation}
\Phi = \Phi_C + \Phi_D + \Phi_H. \end{equation} The central
component of the Galactic potential in cylindrical coordinates
$(r, \theta, z)$ is represented as
\begin{equation}
\Phi_C=-\frac{M_C}{(r^2+z^2+b_C^2)^{1/2}},
\end{equation}
where $M_C$ is the mass and $b_C$ is the scale parameter. The disk
component is
\begin{equation}
\Phi_D=-\frac{M_D}{\{r^2+[a_D+(z^2+b_D^2)^{1/2}]^{1/2}\}^{1/2}},
\end{equation}
where $M_D$ is the mass, $a_D$ and $b_D$ are the scale parameters.
The halo component is
\begin{equation}
\Phi_H=-\frac{M(R)}{R}-\int_R^{100}{{\frac{1}{R^{'}}}{\frac{dM(R^{'})}{dR^{'}}}}dR^{'},
\end{equation}
where
$$
M(R)=\frac{M_H(R/a_H)^{2.02}}{1+(R/a_H)^{1.02}},
$$
Here, $M_H$ is the mass and $a_H$ is the scale parameter. If $R$
is measured in kpc and $M_C,M_D,M_H$ are measured in units of the
Galactic mass $(M_G)$ equal to $2.32\times10^7M_\odot$, then the
Gravitational constant is $G=1$ and the unit of measurement of the
potential $\Phi$, along with the individual components of
(3)--(5), is 100 km$^2$ s$^{-2}$.

All of the Allen--Santillan model parameters adopted here are
given in Table 1.

If the spiral density wave is taken into account (Lin and Shu
1964; Lin et al. 1969), then the following term is added to the
right-hand side of Eq.~(2) (Fernandez et al. 2008):
\begin{equation}
 \Phi_{sp} (R,\theta,t)= A\cos[m(\Omega_p t-\theta)+\chi(R)],
\end{equation}
where
 $$
 A= \frac{(R_0\Omega_0)^2 f_{r0} \tan i}{m},
 $$$$
 \chi(R)=- \frac{m}{\tan i} \ln\biggl(\frac{R}{R_0}\biggr)+\chi_\odot.
 $$
Here, $A$ is the amplitude of the spiral wave potential; $f_{r0}$
is the ratio of the radial component of the perturbation from the
spiral arms to the Galaxy's total attraction; $\Omega_p$ is the
pattern speed of the wave; $m$ is the number of spiral arms; $i$
is the arm pitch angle, $i<0$ for a winding pattern; $\chi$ is the
phase of the radial wave (the arm center then corresponds to
$\chi=0^\circ$); and $\chi_\odot$ is the Sun's phase in the spiral
wave.

The spiral wave parameters are very unreliable (a review of the
problem can be found in Fernandez et al. (2001) and Gerhard
(2010)). The simplest model of a two-armed spiral pattern is
commonly used, although, as analysis of the spatial distribution
of young Galactic objects (young stars, star-forming regions, or
hydrogen clouds) shows, both three- and four-armed patterns are
possible (Russeil 2003; Englmaier et al. 2008; Hou et al. 2009).
More complex models are also known, for example, the kinematic
model by L\'epine et al. (2001) that combines two- and four-armed
spiral patterns rigidly rotating with an angular velocity close
$\Omega_0$. Note also the spiral ring Galactic model (Mel'nik and
Rautiainen 2009) that includes two outer rings elongated
perpendicular and parallel to the bar, an inner ring elongated
parallel to the bar, and two small fragments of spiral arms. As
applied to the Galaxy, the theories of nonstationary spiral waves
with a fairly short stationarity time (several 100 Myr), a
variable rotation rate, and a variable number of arms are also
considered (Sellwood and Binney 2002; Baba et al. 2009). The
currently available data do not yet allow one of the listed models
to be unequivocally chosen. Therefore, here we apply the model of
a stationary spiral pattern with different numbers of spiral arms.

Given the data on the Galactic bar rotation (Debattista et al.
2002), the pattern speed $\Omega_p$ can lie within the range
15--65 km s$^{-1}$ kpc$^{-1}$. The possibility of the coexistence
of several pattern speeds, a rapidly rotating bar and a slower
spiral pattern, is also considered (Minchev and Famaey 2010;
Gerhard 2010). Here, we disregard the influence of the bar.
Therefore, we choose $\Omega_p$ from the range 15--30 km s$^{-1}$
kpc$^{-1}$ (Popova and Loktin 2005; Naoz and Shaviv 2007; Gerhard
2010).

The pitch angle $i$ is known relatively well and is
$-5^\circ\div-7^\circ$ and $-10^\circ\div-14^\circ$ for the two-
and four-armed spiral patterns, respectively.

The amplitudes of the velocities of the perturbation from the
spiral density wave are $5-10$ km s$^{-1}$ (Mishurov and Zenina
1999; Fernandez et al. 2001; Bobylev and Bajkova 2010) in both
tangential and radial directions.

The Sun's phase in the wave $\chi_\odot$ is known with a very
large uncertainty. For example, according to Fernandez et al.
(2001), this angle lies within the range $284^\circ-380^\circ$.
Having analyzed the kinematics of OSCs (Bobylev et al. 2008;
Bobylev and Bajkova 2010), we found $\chi_\odot$ close to
$-117^\circ$ (or $243^\circ$), while the data on masers yielded an
estimate of $\chi_\odot = -130\pm10^\circ$. (Bobylev and Bajkova
2010); here, the minus implies that hat we measure the phase angle
from the center of the Carina--Sagittarius spiral arm. From the
position of the Sun among the spiral arms, it is obvious that this
angle is close to $-\pi/2$ (Russeil 2003).

According to the classical approach in the linear density-wave
theory (Yuan 1969), the ratio $f_{r0}$ lies within the range
0.04--0.07 and the most probable value is $f_{r0} = 0.05$. The
upper limit $f_{r0} = 0.07$ is determined by the velocity
dispersion of young objects observed in the Galaxy (at $f_{r0} =
0.07,$ the dispersion must reach 25~km s$^{-1}$, which exceeds a
typical observed value of 10--15~km s$^{-1}$).

We adopted the following parameters: the two-armed spiral pattern
$(m=2),$ the pattern speed of the spiral wave $\Omega_p=20$ km
s$^{-1}$ kpc$^{-1}$, the pitch angle $i=-5^\circ,$ the ratio
$f_{r0}=0.05,$ and the Sun's phase in the wave $\chi_\odot =
-117^\circ$. We also took into account the Sun's displacement from
the Galactic plane $Z_\odot=17$~pc (Joshi 2007) and used the
present-day peculiar velocity of the Sun relative to the local
standard of rest $(U_\odot,V_\odot,W_\odot)_{LSR}=(10,11,7)$~km
s$^{-1}$ (Binney 2010; Sch\"{o}nrich et al. 2010; Bobylev and
Bajkova 2010).

For all of the selected 162 stars, we determined the relative
distances $d$ between the star and the Sun as well as their
velocity difference $dV$ in the time interval in the past with a
boundary of $-4.5$~Gyr. The minimum values of these parameters,
$d_{min}$ and $dV_{min}$, were also determined for a certain time
$t_{min}$.

\begin{table}
\caption{Data on the stars}
  \label{Tab2}
\begin{center}
\begin{tabular}{|c|c|c|r|c|c|c|c|}\hline
  HIP   &  SP &  Age, Gyr      & [Fe/H]  & Reference & $d_{min},$ pc & $t_{min},$ Gyr & $dV_{min},$ km s$^{-1}$
  \\\hline

 47399  & F8V & 4.3 (  0--7.3) & $-0.21$ & (1)   & 1        &   $-3.5$   & 2           \\
        &     &                & $-0.10$ & (2)   &          &            &             \\\hline
 87382  & F8V & 3.5 (2.6--4.0) & $+0.04$ & (1)   & 4        &   $-3.1$   & 6           \\
        &     & 3.7 (3.2--4.2) & $+0.01$ & (3,2) &          &            &             \\\hline

\end{tabular}
\end{center}
{\small Note. For the age, the upper and lower limits of the
estimate calculated with a 1ó error are given in parentheses;
1---Holmberg et al. (2009), where the metallicity was determined
from Str\"{o}mgren photometry; 2---Robinson et al. (2007), where
the metallicity was determined from Lick indices; 3---Takeda et
al. (2007).}
\end{table}

\begin{figure}[t]{
\begin{center}
 \includegraphics[width=100mm]{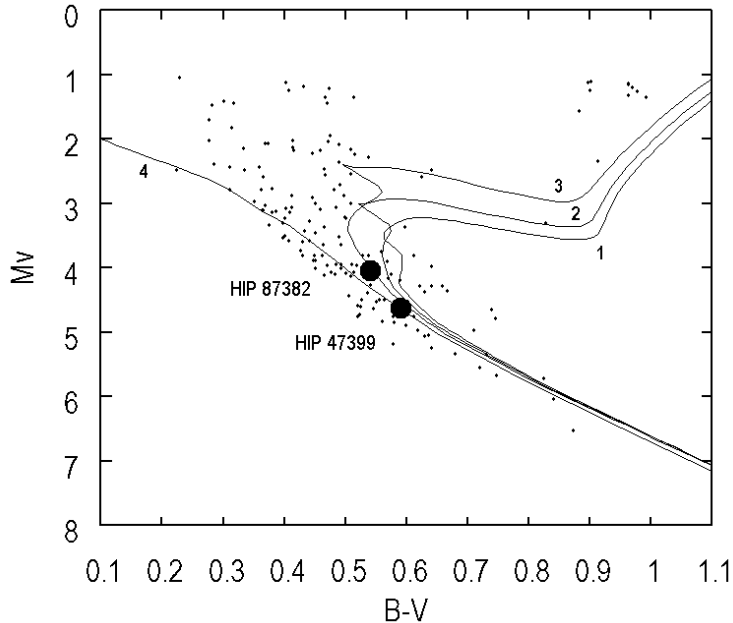}
 \caption{
 Positions of stars on the color–absolute magnitude diagram;
 the zero-age main sequence (4) and three isochrones
 (Demarque et al. 2004) for ages of 3 (3), 4 (2), and 5 (1) Gyr are marked.}
 \end{center}}
 \end{figure}

\begin{figure}[p]{
\begin{center}
  \includegraphics[width=70mm]{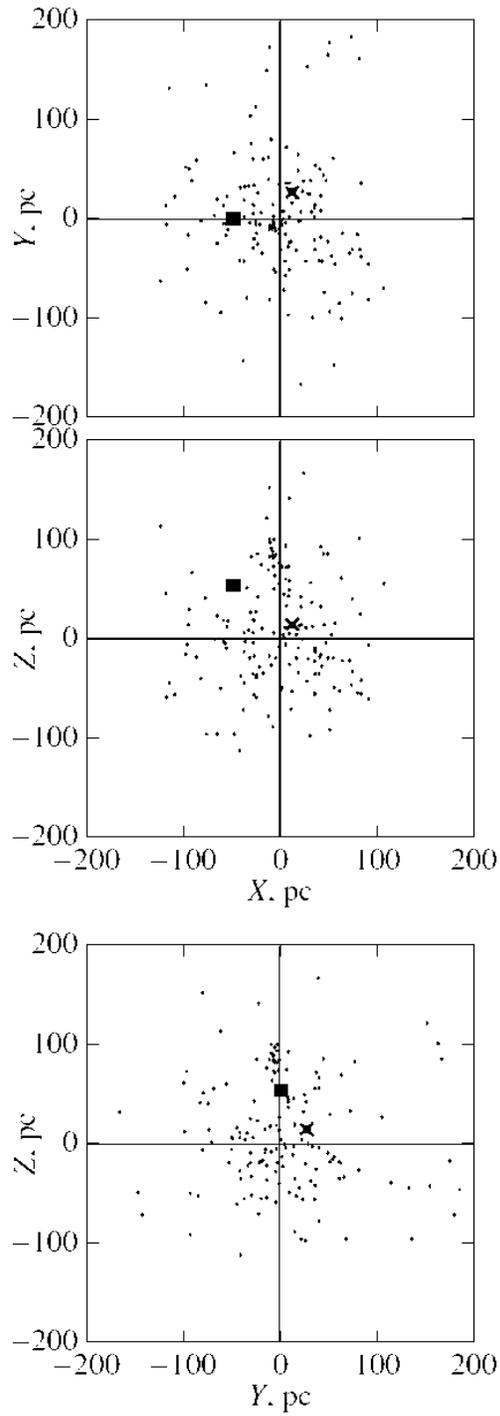}
 \caption{
 Distributions of stars in the $XY,ZX,$ and $YZ$ planes;
 the filled square and cross mark the stars HIP~47399 and HIP~87382, respectively.}
 \end{center}}
 \end{figure}

\section*{RESULTS}
\subsection*{Encounter Parameters}

Figure 1 presents the $(B-V)$ color--absolute magnitude $M_V$
diagram for the 162 selected candidates. To calculate $M_V,$ we
used the apparent $V$ magnitudes from the Hipparcos catalog. We
see from Fig.~1 that the number of candidates lying close to the
isochrones and having nearly solar spectral types will be
considerably smaller than the total number of stars selected by
their kinematics at the first step.

The distributions of the sample of 162 stars in the Galactic
$XY,ZX,$ and $YZ$ coordinate planes are shown in Fig.~2. The
$X,Y,$ and $Z$ coordinate axes are directed toward the Galactic
center, in the direction of Galactic rotation, and toward the
Galactic Pole, respectively.

We found that only two stars from our sample, namely HIP~47399
(Fig.~3) and HIP~87382 (Fig.~4), can be acceptable candidates. For
them, there is good agreement between their estimated ages and the
kinematic encounter parameters $(d,dV)$ found. In Figs.~1 and 2,
the positions of these stars are marked by the large filled
circles (the random error bars are within the circles) and the
crosses, respectively.

For HIP~47399, we have the following initial data:
 $\alpha=9^h 39^m 27^s.4,$
 $\delta = 42^\circ 17' 09'',$
 $\mu_\alpha\cos \delta=-4.29\pm1.38$ mas yr$^{-1}$,
 $\mu_\delta = -6.70\pm0.59$ mas yr$^{-1}$,
 $\pi = 13.87\pm0.95$ mas (the heliocentric distance $r= 72\pm5$~pc),
 the radial velocity
 $V_r =-7.3\pm3.4$ km s$^{-1}$,
 $U = 3.9\pm2.3$ km s$^{-1}$,
 $V = -2.5\pm0.3$ km s$^{-1}$,
 $W = -6.3\pm2.6$ km s$^{-1}$.

For HIP~87382:
 $\alpha= 17^h 51^m 14^s.0,$
 $\delta= 40^\circ 04' 20'',$
 $\mu_\alpha\cos \delta= -16.86\pm0.28$ mas yr$^{-1}$,
 $\mu_\delta = 11.01\pm0.42$ mas yr$^{-1}$,
 $\pi= 29.76\pm0.36$ mas ($r=33.6\pm0.4$~pc),
 $Vr = 1.6\pm0.2$ km s$^{-1}$,
  $U =-1.0\pm0.1$ km s$^{-1}$,
  $V = 0.4\pm0.2$ km s$^{-1}$,
  $W = 3.4\pm0.1$ km s$^{-1}$.
The ages and metallicities of these stars estimated by various
authors are given in Table 2. The error in [Fe/H] is typically
0.1--0.15 dex. This suggests that the metallicity of HIP~47399 and
HIP~87382 is nearly solar, within the error limits.

HIP~21158 and HIP~99689 common to the list by Brown et al. (2010)
do not withstand the described test for close encounters in the
past.

The encounter parameters $d$ and $dV$ depend on the adopted
components of the Sun's peculiar velocity vector
$(U_\odot,V_\odot,W_\odot)_{LSR}$ and on whether or not we take
into account the influence of the spiral wave. Without allowance
for the spiral wave, there are no close encounters between HIP
47399 and the Sun, while the closest encounters for HIP~87382
($d_{min}=16$~pc, $dV_{min}=16$~km s$^{-1}$) occur at
$t_{min}=-1.2$~Gyr. It may be concluded that allowance for the
spiral density wave has a significant influence on the results.

The encounter parameters calculated for HIP~47399 and HIP~87382 by
taking into account the influence of the spiral wave are given in
the last columns of Table~2.

\begin{figure}[t]{
\begin{center}
  \includegraphics[width=100mm]{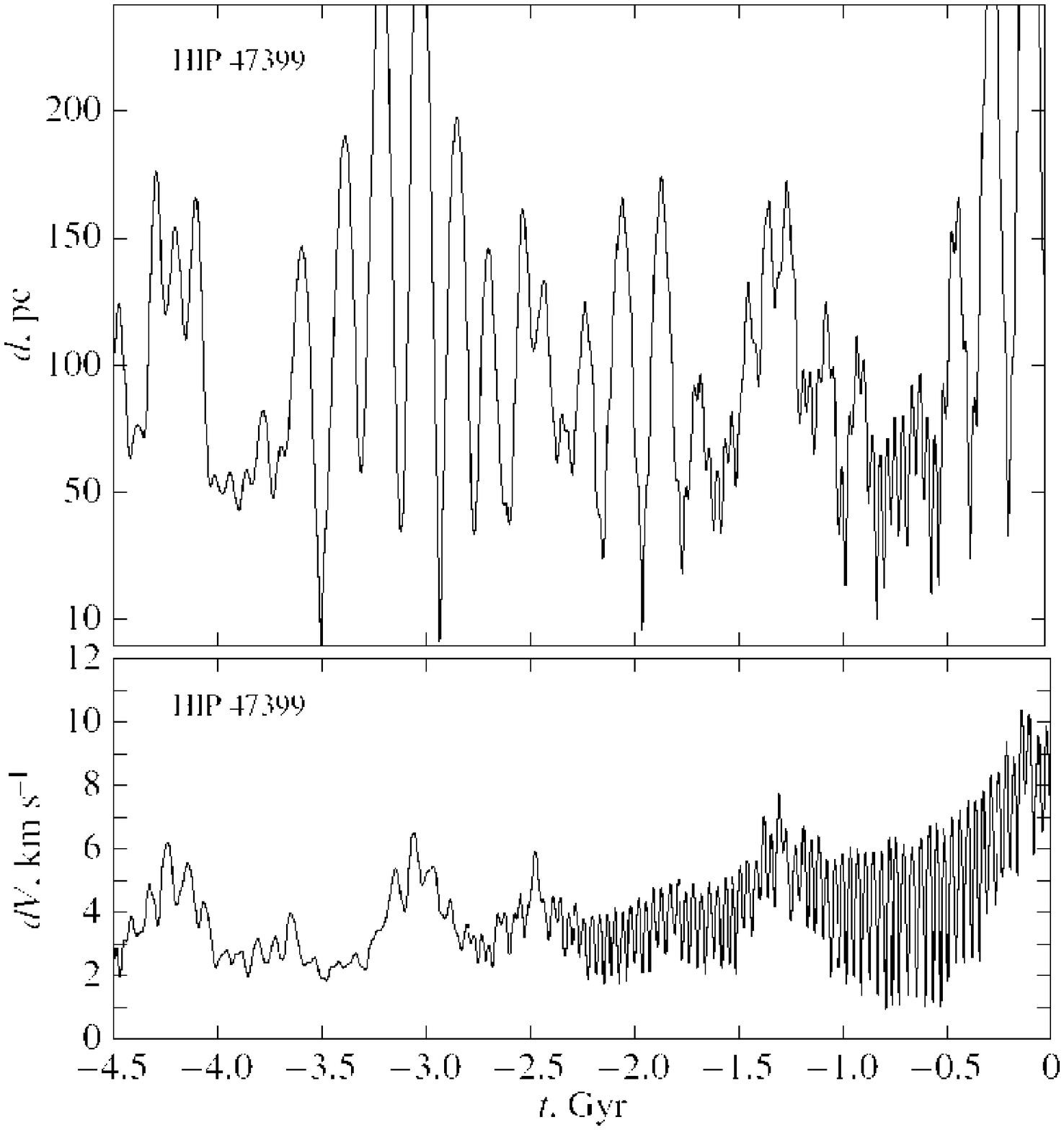}
 \caption{Parameters $d$ and $dV$ of the encounter between
 HIP~47399 and the solar orbit versus time for the two-armed spiral pattern.}
 \end{center}}
 \end{figure}
\begin{figure}[t]{
\begin{center}
  \includegraphics[width=100mm]{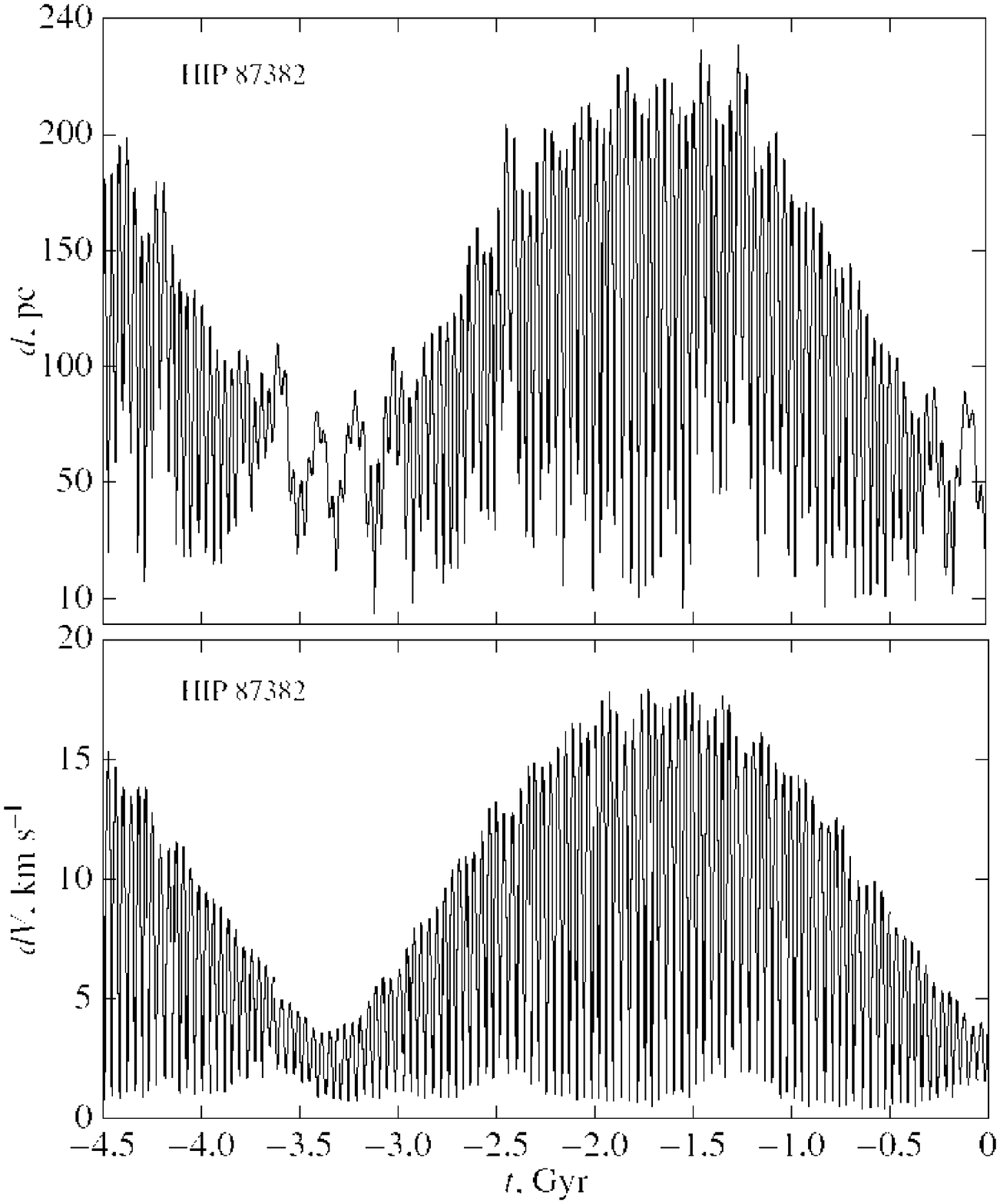}
 \caption{Parameters $d$ and $dV$ of the encounter between
 HIP~87382 and the solar orbit versus time for the two-armed spiral pattern.}
 \end{center}}
 \end{figure}
\begin{figure}[t]{
\begin{center}
  \includegraphics[width=100mm]{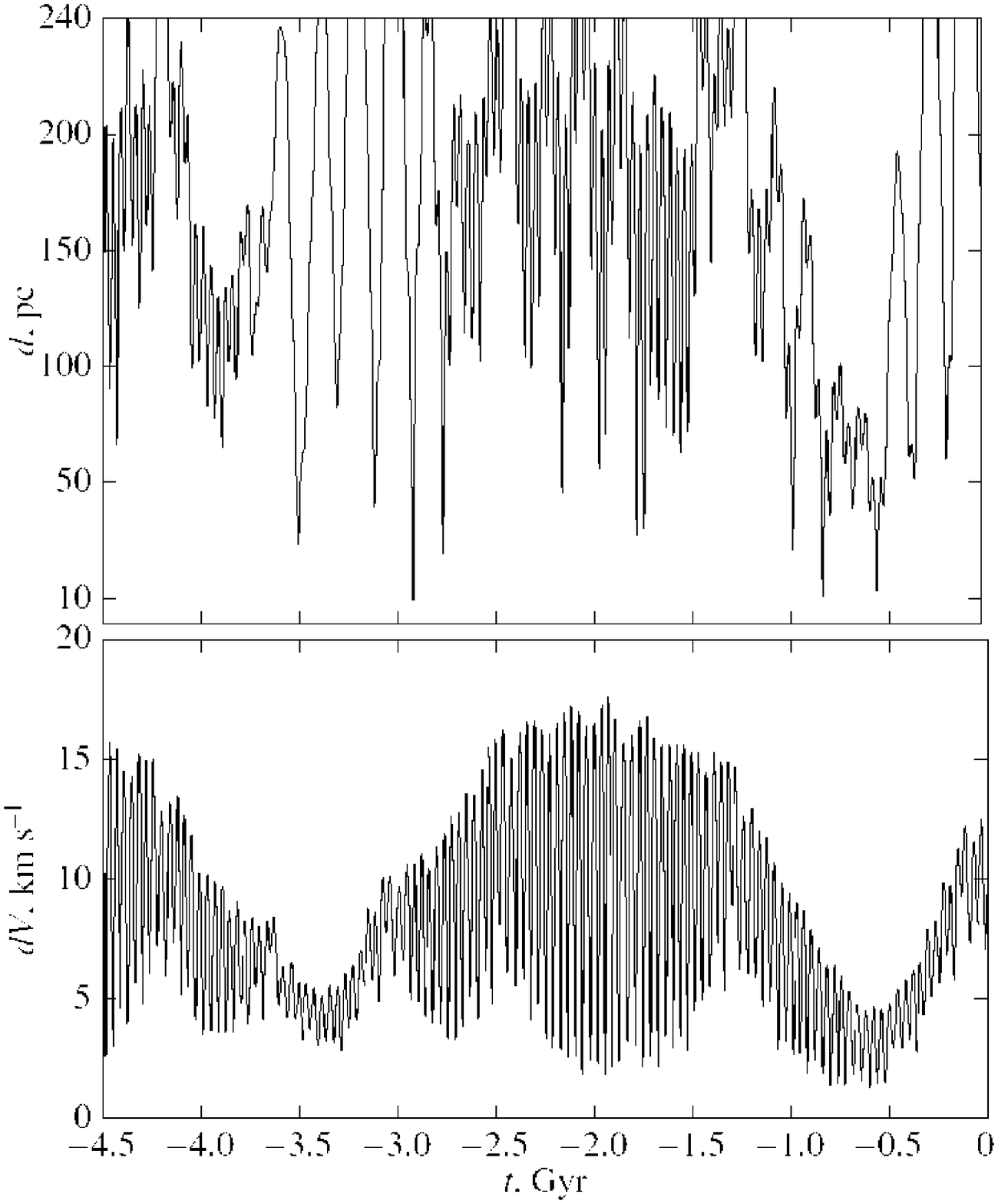}
 \caption{Parameters $d$ and $dV$ of the mutual encounters between
 HIP~47399 and HIP~87382 versus time for the two-armed spiral pattern.}
 \end{center}}
 \end{figure}

\subsection*{The Probability of Close Encounters}

To estimate the probability that a star was born together with the
Sun in a common star cluster, we should take into account the
following. The lifetime of the OSC as a gravitationally bound
structure is typically no more than 2~Gyr. The OSC core size is
typically about 10~pc. The OSC has an open halo and a tail
elongated along its orbit develops with time. We assumed that the
OSC was born at a time of $-4.5$~Gyr.

We then consider the probability $(p)$ that the star and the Sun
belong to the OSC core to be nonzero if the following conditions
are met for them in the time interval $-4.5<t<-2.5$~Gyr: the
relative distance along the Galactic radius is $|dR| < 20$~pc,
along the vertical axis is $|dZ|<20$~pc, and along the orbit is
$|dY|<20$~pc, then $d = \sqrt {dR^2 + dZ^2 + dY^2}<35$~pc. For
HIP~87382, according to the upper panel in Fig.~4, these
conditions are close to fulfilment in this time interval---we
observe a minimum of $d=40-60$~pc near $-3.5$~Gyr. Similar
reasoning is also valid for HIP~47399.

As a result, we calculate $p$ as the ratio of the number of points
in the narrow peaks (Figs.~3 and 4) satisfying the adopted
constraints to the total number of points in a given time
interval.

The constraint on the relative velocity at the encounter time
should also be taken into account. Then, for example, for
HIP~87382 at $d < 10$~pc and $|dV| < 10$ km s$^{-1}$ in the time
interval from $-4.5$ to $-2.5$~Gyr, the probability $p=0.001
(0.1\%)$ is then a more rigorous estimate. For less rigorous
requirements, $d<20$~pc and $|dV|<10$ km s$^{-1}$, $p=0.006
(0.6\%)$.

For HIP~47399, these numbers are slightly larger (because its
relative velocity is always low, $dV < 10$ km s$^{-1}$, as we see
from the lower panel in Fig. 3): $p = 0.006 (0.6\%)$ for $d<10$~pc
and $|dV|<10$ km s$^{-1}$ and $p = 0.014 (1.4\%)$ for $d<20$~pc
and $|dV|<10$ km s$^{-1}$.

For a rigorous estimation of $p$, we should take into account the
errors in the observational data and the errors in the parameters
of the Sun's peculiar velocity and the parameters of the spiral
density wave using the method of statistical simulations. In
Fig.~5, the parameters of the mutual encounters between HIP~87382
and HIP~87382 are plotted against time. We can see that their
relative velocity in the time interval from $-3.5$ to $-2.8$~Gyr
is 5--10 km s$^{-1}$; there is a peak encounter to distances of
less than 10 pc. Hence it may be concluded that within the
framework of the approach used, three objects (the Sun, HIP~47399,
and HIP~87382) are possible candidates for being members of a
common open cluster.

\begin{table}
 \caption{Encounter parameters calculated using the two-armed $(m=2)$ spiral
 pattern at various $f_{r0}$}
    \label{Tab3}
\begin{center}
\begin{tabular}{|c|c|r|c|c|c|c|c|c|}\hline
  HIP   & $f_{r0}$ &   & $d_{min},$ pc & $t_{min},$ Gyr & $dV_{min},$ km s$^{-1}$ \\\hline
 47399  &  0.045   & * & 10       &   $-2.3$   & 3           \\
        &  0.050   &   &  1       &   $-3.5$   & 2           \\
        &  0.055   & * & 11       &   $-2.7$   & 3           \\
        &  0.060   & * & 31       &   $-3.9$   & 9           \\
        &  0.065   &   & 27       &   $-2.9$   & 7           \\\hline
 87382  &  0.045   & * &  6       &   $-3.3$   &  4          \\
        &  0.050   &   &  4       &   $-3.1$   &  6          \\
        &  0.055   &   &  2       &   $-2.1$   & 17          \\
        &  0.060   &   &  5       &   $-2.2$   & 16          \\
        &  0.065   & * & 26       &   $-3.5$   &  4          \\\hline
 \end{tabular}
 \end{center}
 {\small Note. * --- for the local minimum from the interval $t<-2$~Gyr.}
 \end{table}

\subsection*{Statistical Simulations}

We calculated the encounter parameters by taking into account the
random errors. The errors in the Sun's peculiar velocity
components $(U_\odot,V_\odot,W_\odot)_{LSR}$ were taken to be
$(0.5,1,0.3)$ km s$^{-1}$, a $10\%$ level in the initial
velocities $(U,V,W)$ and coordinates $(X,Y,Z)$ of the stars as
well as in the parameters of the spiral density wave.

We performed our simulations for HIP~87382, because it has very
small random errors in the initial coordinates and velocities. It
turned out that there are interesting results in our simulations
even in a symmetric potential (mainly due to the variations in the
Sun's peculiar velocity). Therefore, we present the results for
HIP~87382 in the time interval from $-4.5$ to $-2.5$~Gyr for two
cases: (1) the orbits were constructed only in a symmetric
potential and (2) with the influence of the spiral density wave
added:

1. For $d<10$~pc, $|dV|<10$~km s$^{-1}$, $p=0.0002 (0.02\%),$ à
for $d<10$~pc, $|dV|<5$ km s$^{-1}$, $p=0.0001 (0.01\%).$

2. For $d<10$~pc, $|dV|<10$~km s$^{-1}$, $p=0.000009 (0.0009\%),$
à for $d<10$~pc, $|dV|<5$~km s$^{-1}$, $p=0.000006 (0.0006\%).$

We see that $p$ decreased by less than one order of magnitude in
case 1 and by two orders of magnitude in case 2 compared to the
analogous results described in the preceding section for this
star. Obviously, the uncertainty in the parameters of the spiral
density wave has a decisive influence.

One of the most important parameters in our model of the spiral
wave is the ratio of the radial component of the perturbation from
the spiral arms to the Galaxy's total attraction, $f_{r0}$. To
investigate the robustness (stability to small perturbations) of
our results, we determined the encounter parameters for $f_{r0}$
from the range $0.04-0.07$ (with all of the remaining model
parameters being fixed). The results are presented in Table~3. We
see from this table that the results remain acceptable for our
problem up to $f_{r0} = 0.065$ (the perturbation is $30\%$ with
respect to 0.05).

\begin{figure}[t]{
\begin{center}
  \includegraphics[width=100mm]{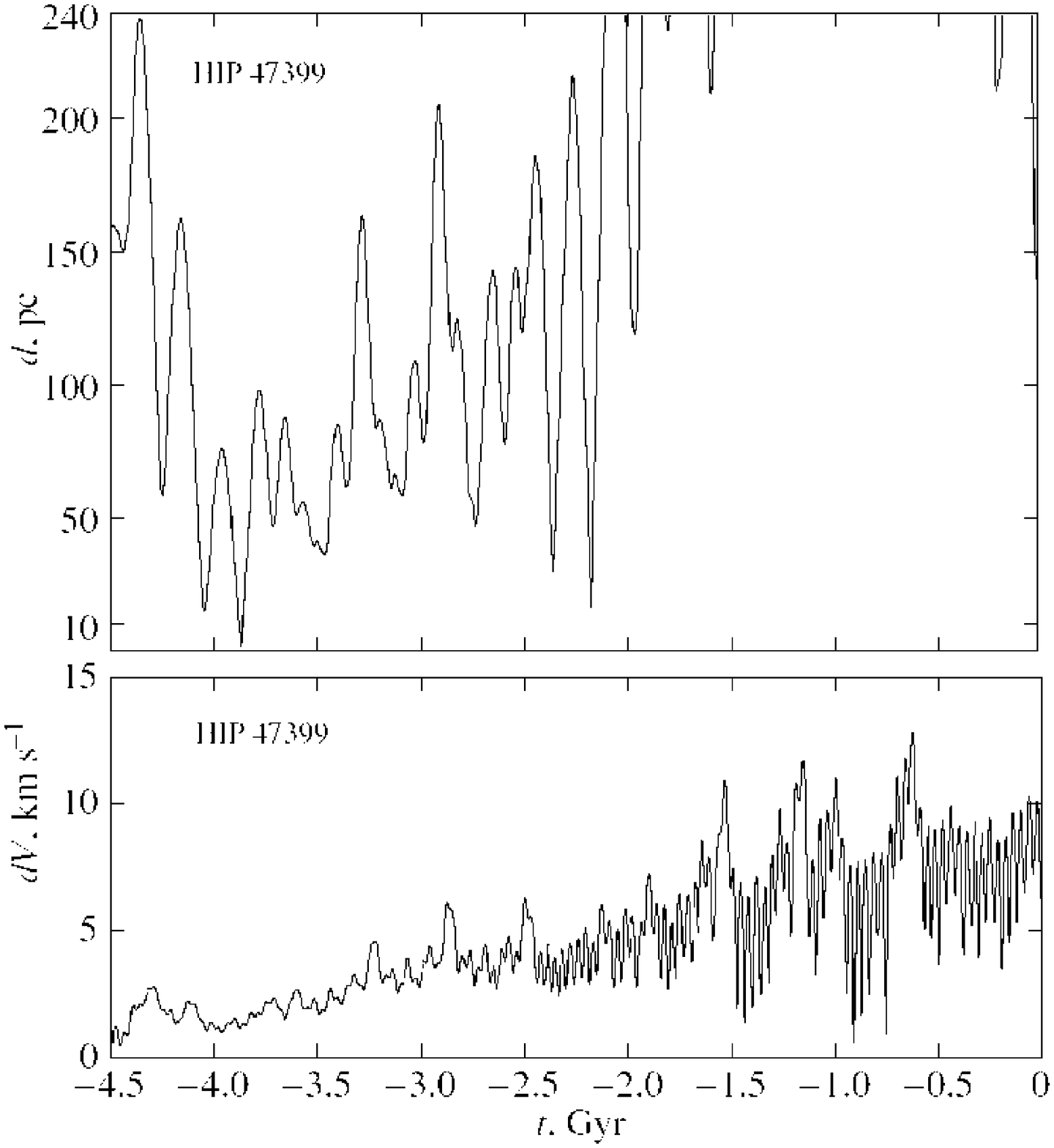}
 \caption{Parameters $d$ and $dV$ of the encounter between
 HIP~47399 and the solar orbit versus time for the four-armed spiral pattern.}
 \end{center}}
 \end{figure}
\begin{figure}[t]{
\begin{center}
  \includegraphics[width=100mm]{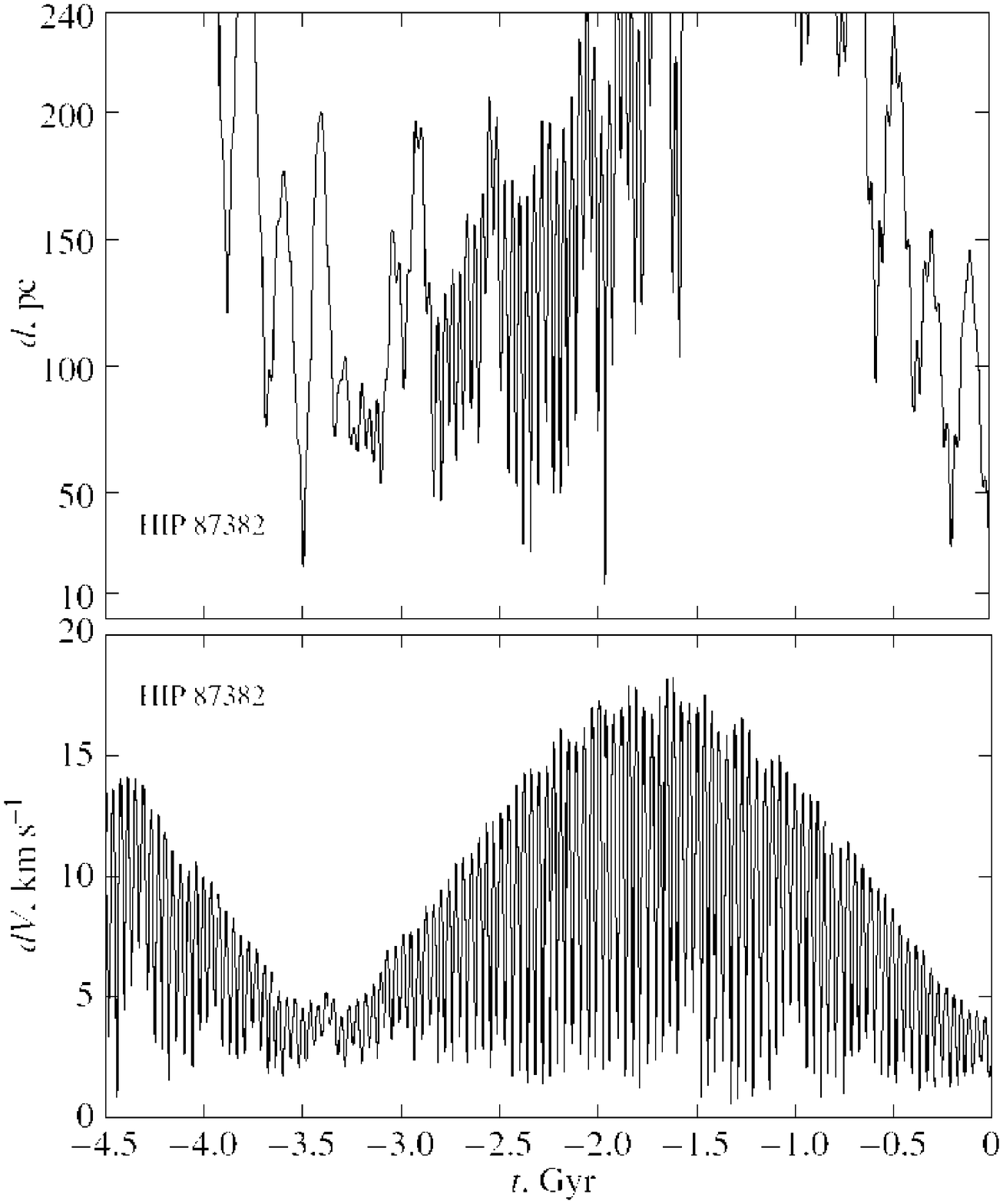}
 \caption{Parameters $d$ and $dV$ of the encounter between
 HIP~87382 and the solar orbit versus time for the four-armed spiral pattern.}
 \end{center}}
 \end{figure}

\subsection*{Other Spiral-Structure Models}

{\bf The four-armed spiral pattern.} For the four-armed spiral
pattern, we set $m=4$ and the pitch angle $i=-10^\circ$. The
remaining parameters were the same as those for the two-armed
model, namely $\Omega_p = 20$ km s$^{-1}$ kpc$^{-1}$,
$f_{r0}=0.05,$ and $\chi_\odot=-117^\circ$. The simulation results
are presented in Figs.~6 and 7.

We can seen from comparison of Figs.~6 and 3 that the four-armed
model of the spiral pattern for HIP 47399 yields more interesting
results: at $t\approx4$~Gyr, a global minimum is observed in
encounters to distances of less than 30 pc with relative encounter
velocities $<2$ km s$^{-1}$.

As can be seen from comparison of Figs.~7 and 4, the picture of
encounters with the solar orbit did not change qualitatively for
HIP 87382.

{\bf The 2+4 spiral pattern.} For the composite $(2+4)$ model of
the spiral pattern (L\'epine et al. 2001; Mishurov and Acharova
2010), the spiral wave potential (6) contains two terms with
amplitudes $A_2$ (two-armed component) and $A_4$ (four-armed
component). In this model, the Sun is very close to the corotation
circle and, as Mishurov and Acharova (2010) showed, the influence
of the spiral structure is so strong that the test model particles
are scattered over a significant spatial volume.

The following model parameters were adopted. For the two-armed
component: $m=2, i_2=-7^\circ, \chi_\odot=300^\circ$. For the
four-armed component: $m=4, i_4=-14^\circ, \chi_\odot=135^\circ$.
Both spiral patterns rotate with the same angular velocity
$\Omega_p=\Omega_0.$ In our case, $\Omega_0=220/8.5=25.9$ km
s$^{-1}$ kpc$^{-1}$, as follows from the Allen--Santillan (1991)
model parameters. We took $f_{r0}=0.05$ when calculating $A_2$ and
used $A_2/A_4=0.8$ to determine $A_4.$

The results are presented in Figs.~8 and 9. We can see that for
both stars, HIP~47399 and HIP~87382, there are encounters of
interest to us.

\begin{figure}[p]{
\begin{center}
  \includegraphics[width=100mm]{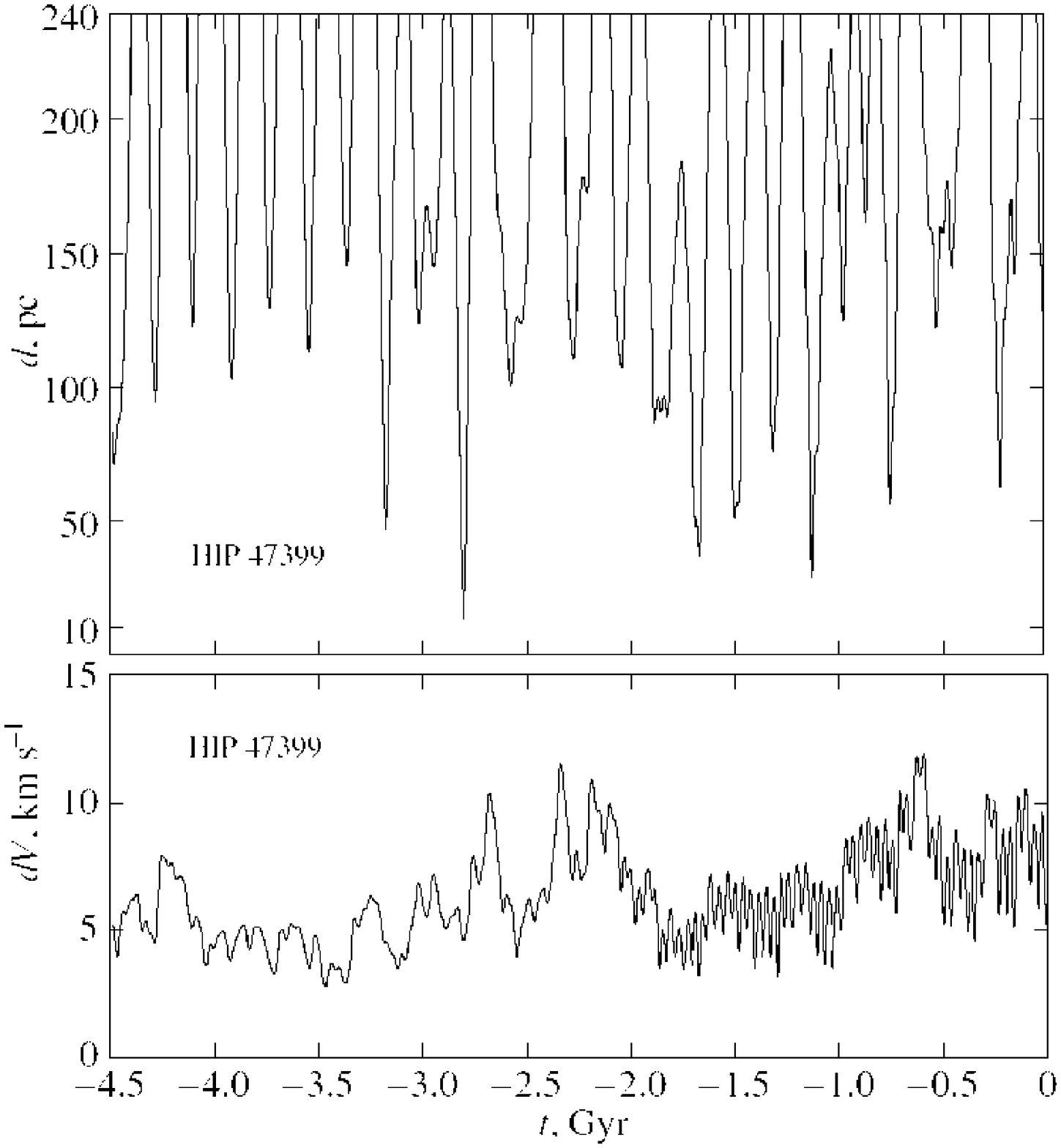}
 \caption{Parameters $d$ and $dV$ of the encounter between
 HIP~47399 and the solar orbit versus time for the composite $(2+4)$ spiral pattern.}
 \end{center}}
 \end{figure}
\begin{figure}[p]{
\begin{center}
  \includegraphics[width=100mm]{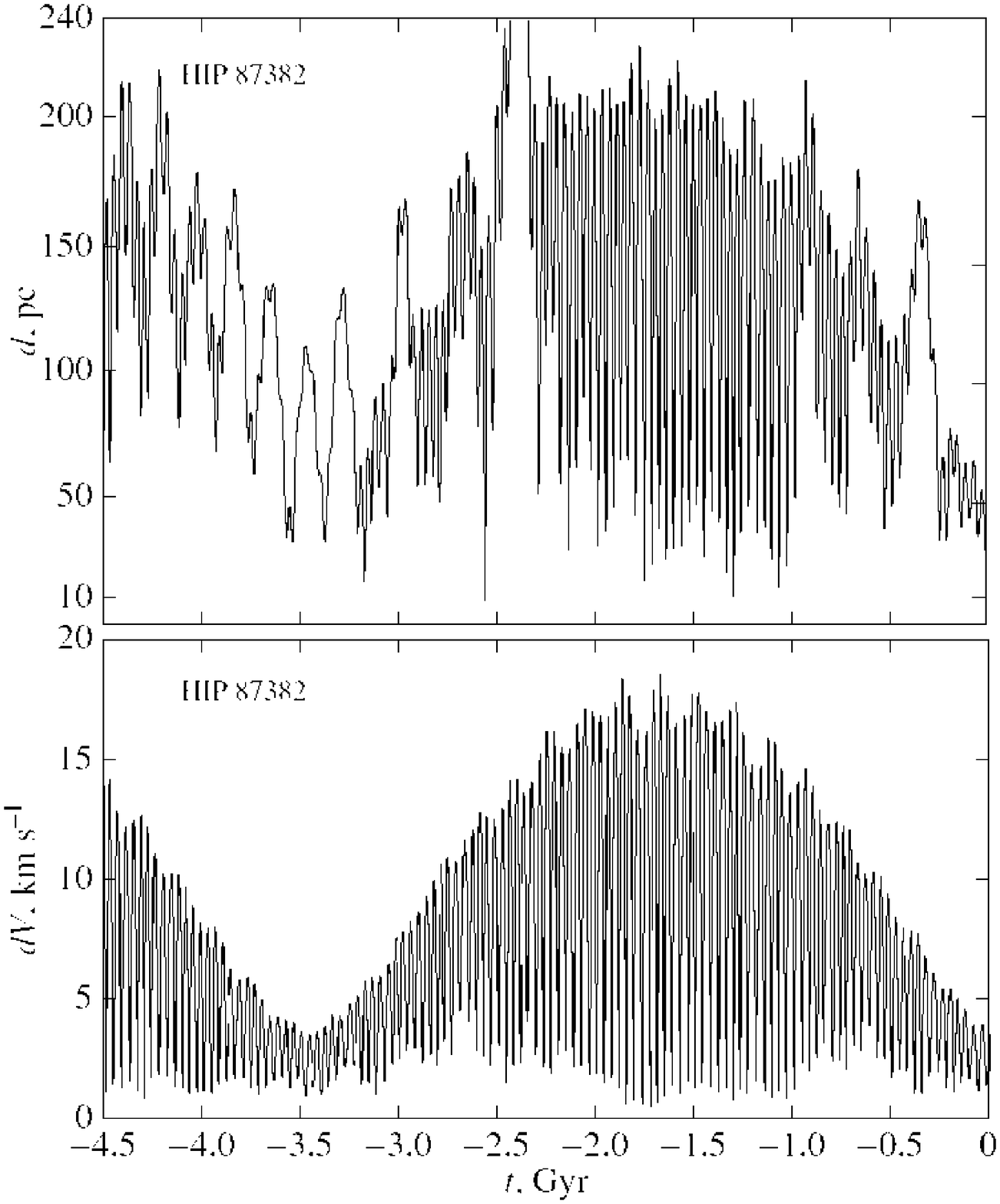}
 \caption{Parameters $d$ and $dV$ of the encounter between
 HIP~87382 and the solar orbit versus time for the composite $(2+4)$ spiral pattern.}
 \end{center}}
 \end{figure}

\section*{CONCLUSIONS}

Based on the proposed kinematic approach to searching for the
Sun's siblings from a common cluster, we selected 162 F, G, and K
stars from the Hipparcos catalog with low heliocentric space
velocities ($<8$ km s$^{-1}$) from a solar neighborhood
$\approx$200 pc in radius. For all these stars, we constructed the
Galactic orbits for 4.5 Gyr into the past using the
Allen--Santillan (1991) axisymmetric model Galactic potential that
additionally included the perturbations from the spiral density
wave. The parameters of the encounter with the solar orbit were
calculated for each orbit.

We confirmed the conclusion reached by Mishurov and Acharova
(2010) that the spiral density waves have a significant influence
on the search results.

We considered the influence of a stationary spiral structure for
the two- and four-armed models $m=2,4$ as well as the composite
$2+4$ model (L\'epine et al. 2001) and found that they all confirm
the results of our search.

We found that almost all of the stars considered in the past
receded from the solar orbit fairly rapidly and far. Two single
stars, HIP 87382 and HIP 47399, constitute an exception. In the
two-armed model of the Galactic spiral pattern, their orbits
oscillate near the solar orbit with an amplitude of $\approx$250
pc; there are peak close encounters to distances $d<10$~pc; both
stars have a wide minimum up to distances $d<60$~pc in the time
interval $-4\div-3$ Myr consistent with their age estimates. HIP
47399 is interesting in that its heliocentric velocity is always
low, while this velocity in the time interval $-4\div-3$ Myr is
$|dV|<3$ km s$^{-1}$. The closest and longest encounters at
$t\approx-4$ Gyr are observed for the four-armed model $(m=4)$ for
HIP 47399 (Fig. 6).

As a result, we conclude that HIP~47399 and HIP~87382 are of
considerable interest as possible candidates for the Sun's
siblings from the hypothetical parent cluster.

\bigskip
{\bf ACKNOWLEDGMENTS}
\bigskip

We are grateful to the referees for helpful remarks that
contributed to an improvement of the paper. We separately thank
Yu.~N. Mishurov for the discussion of the problems considered
here. The SIMBAD search database provided a great help to our
study. This work was supported by the Russian Foundation for Basic
Research (project no. 08--02--0040) and in part by the ``Origin
and Evolution of Stars and Galaxies'' Program of the Presidium of
the Russian Academy of Sciences and the Program of State Support
for Leading Scientific Schools of the Russian Federation (project.
NSh--3645.2010.2, ``Multiwavelength Astrophysical Studies'').

\bigskip
{\bf REFERENCES}
\bigskip

{\small
 1. I.A. Acharova, J.R.D. L\'epine, Yu.N. Mishurov, et al., MNRAS 402, 1149 (2010).

 2. C. Allen and A. Santillan, Rev. Mex. Astron. Astrofis. 22, 255 (1991).

 3. J. Baba, Y. Asaki, J. Makino, et al., Astrophys. J. 706, 471 (2009).

 4. J.J. Binney, MNRAS 401, 2318 (2010).

 5. J. Bland-Hawthorn and K. Freeman, Publ. Astron. Soc. Austral. 21, 110 (2004).

 6. J. Bland-Hawthorn, M.R. Krumholz, and K. Freeman, Astrophys. J. 713, 166 (2010).

 7. V.V. Bobylev, A.T. Bajkova, and A.S. Stepanishchev, Astron. Lett. 34, 515 (2008).

 8. V.V. Bobylev, A.T. Bajkova, and A.A. Myll\"{a}ri, Astron. Lett. 36, 27 (2010).

 9. V.V.Bobylev, and A.T. Bajkova, MNRAS 408, 1788 (2010).

 10. A.G.A. Brown, S.F. Portegies Zwart, and J. Bean, MNRAS 407, 458 (2010).

 11. Ya.O. Chumak, A.S. Rastorguev, and S.D. Aarseth, Astron. Lett. 31, 308 (2005).

 12. Ya.O. Chumak and A.S. Rastorguev, Astron. Lett. 37, 157 (2011).

 13. Ya.O. Chumak and A.S. Rastorguev, Astron. Lett. 34, 446 (2008).

 14. V.P. Debattista, O. Gerhard, and M. N. Sevenster, MNRAS 334, 355 (2002).

 15. P. Demarque, J.Woo, Y.Kim, and S.K. Yi, Astrophys. J. Suppl. Ser. 155, 667 (2004).

 16. P. Englmaier, M. Pohl, and N. Bissantz, arXiv astroph: 0812.3491 (2008).

 17. D. Fernandez, F. Figueras, and J. Torra, Astron. Astrophys. 372, 833 (2001).

 18. D. Fernandez, F. Figueras, and J. Torra, Astron. Astrophys. 480, 735 (2008).

 19. O. Gerhard, Memorie della Societa Astronomica Italiana Supplement, v.18, p.185 (2011).

 20. G.A. Gontcharov, Astron. Lett. 32, 759 (2006).

 21. J. Holmberg, B. Nordstr\"{o}m, and J. Andersen, Astron. Astrophys. 501, 941 (2009).

 22. L.G. Hou, J.L. Han, and W.B. Shi, Astron. Astrophys. 499, 473 (2009).

 23. Y.C. Joshi, MNRAS 378, 768 (2007).

 24. A.H.W. K\"{u}pper, A. Macleod, and D.C. Heggie, MNRAS 387, 1248 (2008).

 25. F. van Leeuwen, Astron. Astrophys. 474, 653 (2007).

 26. J.R.D. L\'epine, Yu.N. Mishurov, and S.Yu. Dedikov, Astrophys. J. 546, 234 (2001).

 27. C.C. Lin and F.H. Shu, Astrophys. J. 140, 646 (1964).

 28. C.C. Lin, C. Yuan, and F.H. Shu, Astrophys. J. 155, 721 (1969).

 29. A.M. Mel'nik and P. Rautiainen, Astron. Lett. 35, 609 (2009).

 30. I. Minchev and B. Famaey, Astrophys. J. 722, 112 (2010).

 31. Yu.N. Mishurov and I.A. Zenina, Astron. Astrophys. 341, 81 (1999).

 32. Yu.N. Mishurov and I.A. Acharova, MNRAS 412, 1771 (2011).

 33. S. Naoz and N. J. Shaviv, New Astron. 12, 410 (2007).

 34. L. Pomp\'eia, T. Masseron, B. Famaey, et al., arXiv astro-ph:1101.2583 (2011).

 35. M.E. Popova and A.V. Loktin, Astron. Lett. 31, 171 (2005).

 36. S.F. Portegies Zwart, Astrophys. J. 696, L13 (2009).

 37. S.E. Robinson, S.M. Ammons, K.A. Kretke, et al., Astrophys. J. Suppl. Ser. 169, 430 (2007).

 38. D. Russeil, Astron. Astrophys. 397, 134 (2003).

 39. R. Sch\"{o}nrich, J. Binney, and W. Dehnen, MNRAS 403, 1829 (2010).

 40. J.A. Sellwood and J.J. Binney, MNRAS 336, 785 (2002).

 41. I.I. Shevchenko, Astrophys. J. 733, 39 (2011).

 42. G. Takeda, E.B. Ford, A. Sills, et al., Astrophys. J. Suppl. Ser. 168, 297 (2007).

 43. The HIPPARCOS and Tycho Catalogues, ESA SP--1200 (1997).

 44. M. Valtonen, P. Nurmi, J.-Q. Zheng, et al., Astrophys. J. 690, 210 (2009).

 45. R. Wielen, B. Fuchs, and C. Dettbarn, Astron. Astrophys. 314, 438 (1996).

 46. J. Williams, Contemp. Phys. 51, 381 (2010).

 47. C. Yuan, Astrophys. J. 158, 889 (1969).


}

\end{document}